\documentclass{article}
\usepackage{sao2,psfig}
\begin{document}
\setcounter{page}{57}
\issue{2003, 56, 57-90  }
\markboth{Trushkin, Bursov, Nizhelskij }{The multifrequency monitoring of microquasars. SS433}
\title{The multifrequency monitoring of microquasars. SS433}
\author{S. A. Trushkin \and N. N. Bursov \and N. A. Nizhelskij}
\institute{\saoname }
\date{December 19, 2003}{December 24, 2003}
\maketitle
\begin{abstract}
The principal results of daily observations with the RATAN-600 radio telescope
of X-ray binary with relativistic jets microquasar SS433  in 1986--2003
are presented. We have measured the flux densities  at 0.96, 2.3, 3.9, 7.7,
11.2 and 21.7 GHz in different sets, duration from a week to some months.
In general there are 940 observations of SS433 and more than 4500 flux density
measurements in the period. Observations show that radio spectra
are well fitting by a power law. The mean spectral index remained the same,
$-0.60\pm0.14$ during almost 20 years at least, and mean accuracy of the index
determination was better than 0.1 in our multi-frequency observations,  i.e.
it was higher than in the intensive two-frequency monitoring of
SS433  with the three-element GBI interferometer.
Flux density data and spectra `on-line' plotting   are accessible on the CATS
data base site: http://cats.sao.ru/.
\end{abstract}

\section{Introduction}

The X-ray binaries (XBs) have long been studied in radio band. After an
identification of Cyg X-3 in the beginning of the 70-s it became evident
that XBs could have powerful variable radio emission. A sample of X-ray
binaries with relativistic jets that Mirabel {\&} Rodriguez (1999) named
microquasars consists of 15--20 objects. The brightest of them were observed
actively under a monitoring program of XBs with the RATAN-600 radio telescope
(Trushkin 2000, Trushkin \& Bursov 2001).

SS433 --- a bright variable emission star
--- was identified by Clark and Murdin (1978) with a
rather bright compact radio source 1909+048 located in the center of a 
supernova remnant W50. When in 1979 mobile optical emission lines were 
discovered in the spectrum of this bright star SS433 --- a radio source
1909+04 (Margon et al. 1979), it became apparent that a new class
of objects in the Galaxy was found. At the same time Spencer (1979) was the first to
discover an extended structure: a compact core and 1 arcsec long aligned jets
in the radio image of SS433. At present such a structure in microquasars is commonly
named a radio jet. Different data do indicate a presence of a very narrow
(about 1$\degr$) collimated beam at
least in X-ray and optical ranges. At present there is no doubt that SS433 
is related to W50. SS433 is probably a stellar remnant of
a SN that exploded in a binary system about 30000~years ago. An explosion of
one of the components did not destroy the binary. A distance to SS433 of 4.8
kpc was later determined by different ways including the direct measurement of
proper motions of the jet radio components.

As is now known, this unique variable X-ray, IR and radio source is an 
eclipsing binary cosisting of a compact object and an early type massive star. The
system has a pair of opposite relativistic (in which matter is ejected at
0.26c) jets, which show themselves in
different ways in different ranges from X-ray to radio. The VLBI (Very Long
Baseline Radio Interferometry) observations of the jets proper motion showed
that they consist of separate blobs that move ballistically at the same
velocity  0.26c. Resultant from the 164-day precession of a thick accreting
disk the jets rotate along conic surface and look like structures
similar to a twin corkscrew on VLA radio maps (Hjellming {\&} Johnston 1981).

There are strong evidences for believing that this binary system consists of a
relativistic star with an accreting disk and a massive optical component 
filling its Roche lobe. The supercritical accretion onto the relativistic star
is likely to lead to initiation of the SS433 main feature --- two opposite
jets of matter moving from the accretion disk poles at a velocity of about a 
quarter of the speed of light. The recent intensive spectral
investigation made it possible to determine rather precisely the SS433 mass
function and to estimate the relativistic component mass. It is equal to 
11~$\pm $~5~M$_{\sun}$ (Gies et al. 2002a,b) what suggests that this is
a black hole. Absorption lines in the SS433 spectrum belonging to the
optical component are almost identical with those in the spectrum of an evolved
star of a mid-A spectral class with a mass of 19~$\pm $~7~M$_{\sun}$.

These jets that had been first discovered in spectral optical
observations are seen also in X-ray and radio wavelengths. Activity phenomena
in SS433 give rise to strong variability of all its electromagnetic spectrum.

As is seen from the jet X-ray images and from an oblong form of the radio 
remnant W50, the jet structure extends up to size scales of $1\degr$. It
gives an estimate of jets age as $\sim$ 1000 years. A  simple
kinematic model of SS433 (Abell {\&} Margon 1979) with collimated jets was verified
as a whole by many observational data (Margon 1984).

SS433 is still the unique object in the Galaxy, in which the relativistic
opposing jets
appear in Doppler-shifted optical and X-ray emissions, i.e. beside the 
relativistic radio-emitting electron-positron plasma a baryonic matter
with high atomic numbers moves in the SS433 jets (Kotani et al. 1996, 
Marshall et al. 2002). 

\begin{figure*}
\centerline{\vbox{
\psfig{figure=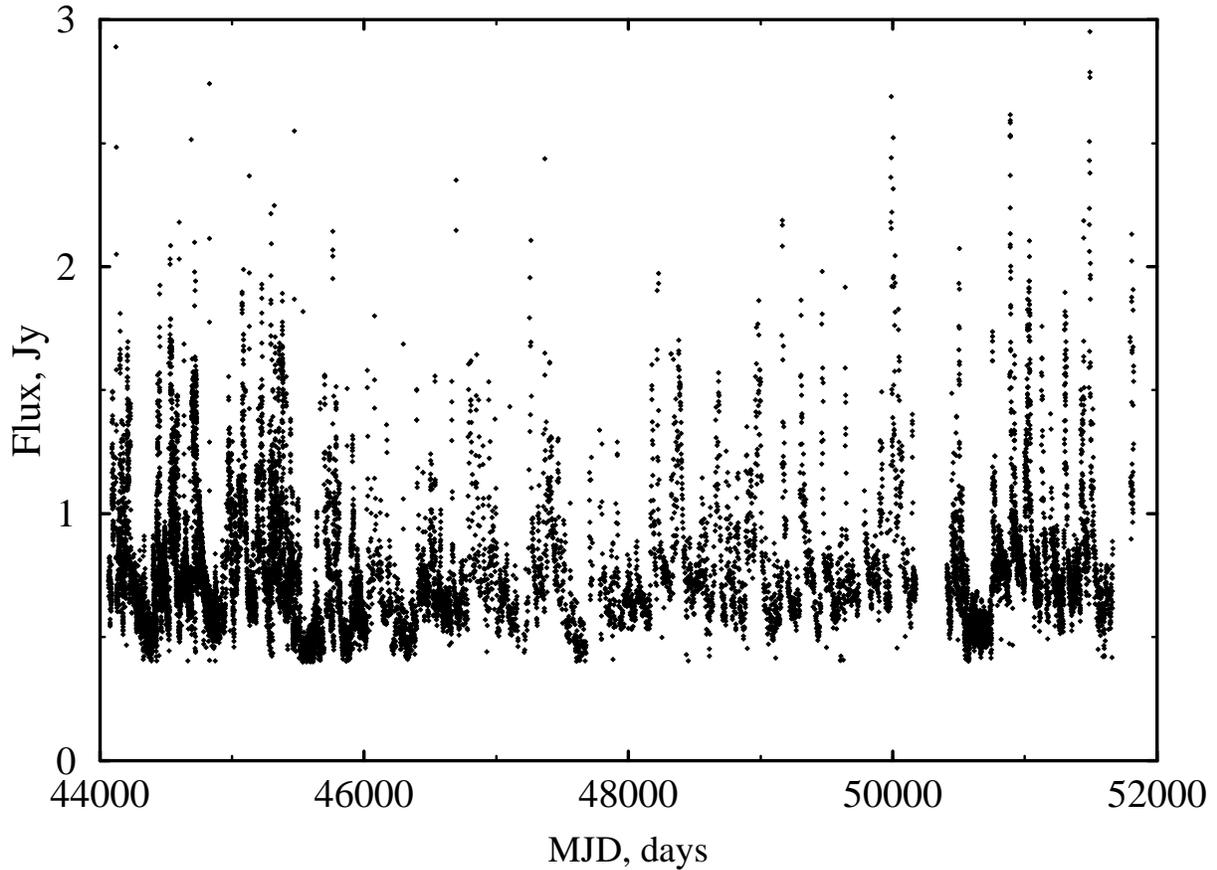,width=16cm,angle=-90}
}}
\caption{%
The SS433 light curves at a frequency of 2.25 GHz by data of
GBI (NRAO/NASA).}
\label{gbi}
\end{figure*}

The major contribution to the study of temporal and spectral properties of
microquasars was made by long-term  program of variable sources
monitoring at two
frequencies of 2.25 and 8.3 GHz with GBI (Green-Bank Interferometer, NRAO).
The publicly available data totally include about 16000
measurements of flux density of SS433. Fig.\ref{gbi} shows the  light curve
of SS433 in 1979--2000 which cover all its GBI-observations at a frequency of 2.25~GHz.
The results of these studies were partly published by Johnston et al.
(1984) and Fiedler et al. (1987).

From recent interesting results of the SS433 study noteworthy is the
circular polarization of its emission in cm range detected by Fender et al.
(2000) and a sign change of the Stokes parameter V detected by
McCormick et al. (2003) in the 1--9~GHz range.
It might be related either to the presence of a gyro-synchrotron
radiation emitted by relatively low-energy electrons or to the change in a
configuration of collimated magnetic field in the jets.

Space observatory INTEGRAL first registered a variable hard X-ray
emission (25--100~keV)
(Cherepashchuk et al. 2003). It was shown that the emission level is
influenced by the precession and orbital motions. These observations in high
energy range confirm indirectly once more a presence of a black hole in this XB.

In radio band a quasisteady synchrotron emission of SS433 with spectrum
S$_{\nu }$~$\sim $~S$_{0}\nu ^{\alpha }$ is superimposed by a non-thermal
flares, when the flux density can exceed the quiescent level 10 times.
The SS433 radio flares are not rare, although the periods of relative
quiescence can be as long as 100--200~days.

At the flares onset there are evidences of synchrotron source opacity
(a flat spectrum) and with further gradual decay of the flare
the spectrum becomes steeper. The
maximum flux of a flare comes later at lower frequencies as was shown for SS433
by Fiedler et al. (1987),
Vermeulen et al. (1993a,b), Bursov {\&} Trushkin (1995); for Cyg X-3 see Waltman
et al. (1994, 1995, 1996) and for GRS~1915+105 --- Trushkin et al. (2001) and
Fender et al. (2002). It corresponds to evolution of an
adiabatically expanding cloud of relativistic emitting electrons. Shklovskij
(1960) and van der Laan (1966) were the first to formulate basic equations
describing this evolution. 

Different modifications of this model necessary for observations fitting
include a hollow conic geometry of jets, an account for synchrotron losses
 or reverse Compton scattering, a dynamic motion
outwards a thermally absorbing envelope. Computations of flares  based on  a
model of twin conic jet show a remarkable coincidence with observations
for Cir\,X-1 (Garcia 1995),   SS433 (Hjellming {\&} Johnston 1988),
LSI+61$\degr$\,303
(Paredes et al. 1991), {\mbox Cyg\,X-3} (Marscher at al. 1975, Marti et al. 1992)
and SAXJ1819-254 (Hjellming et al. 2000).

Based on data of ten SS433 flares in 1986--87, Trushkin (1989) showed that
on average they have close frequency power relations for the maximum
flare flux and the time of this maximum occurrence:
$$\Delta S_{m}{\rm(Jy)}=1.3 \nu^{-0.4\pm 0.15},$$
$$\Delta t_{m}{\rm (days)}=5\nu^{-0.4\pm 0.1},$$
where $\nu$ is in GHz.

A gradual flare decay follows one of two time laws. An original model of
Shklovskij -- van der Laan provides the power law. In many microquasars 
there was actually detected the radio emission decay according to the law 
$$S_\nu = S_o\,t^{-2p}\nu^{\alpha},$$ where $p\approx 2$
is an index of a power electron distribution that is related to the spectral 
index $$\alpha = (1-p)/2\approx-0.5$$ for an optically thin source.
It is in excellent agreement with a possible diffusive mechanism
of relativistic electrons acceleration in shock waves. At a compression 
index in a strong shock wave $\rho\le4$ the spectral index of these
electrons is $$p = (\rho+2)/(\rho-1) = 2.$$

\begin{table*}
\caption { A flux density sensitivity of the ``Northern Sector'' of RATAN-600}
\begin{center}
\medskip
\begin{tabular}{l|l|l|l|l|l|l|l}
\noalign{\smallskip}
\hline
  Wavelength (cm) &31.2  &13.0 & ~7.6  &~6.2  & ~3.9 &  ~2.7& 1.38 \\
\hline
  Frequency (GHz)    & 0.96 &~2.3 & ~3.9  &~4.9  & ~7.7 &  11.2& 21.7  \\
\hline
  $\Delta$S(mJy)   & 40   & 15  &   ~3  &  ~3  &   8  &   10 &  20   \\
\hline
\end{tabular}
\end{center}
\end{table*}

Other flares and fluxes of plasmons in jets were decaying according to the 
exponential law $$S_{\nu }=S_{0}e^{-t/ \tau }$$ (for SS433
see Jowett \& Spencer 1995). A diminution of the characteristic time
$\tau $ with frequency was often but not always observed.
A relation $$\tau {\rm(days)}=11.5~\nu ^{-0.29\pm 0.03}$$ where $\nu$ is in GHz was
observed  in Cyg X-3 (Trushkin 1998). In Cyg X-3 a transit from the exponential law
to the power one was detected several times (Hjellming et al. 1974, Marscher et
al. 1975, Bursov {\&} Trushkin 1995, Trushkin 1998) what is interpreted
as a transit from a mode with a dominated radiative losses to a mode of
adiabatic expansion.

Spencer (1996) calculated the energy of the jet components based on the
condition for minimum energy of relativistic particles and of magnetic field
what is fulfilled at an equipartition of energy of these components.
He showed that a considerable not to say prevailing part
of energy released during flares belonged to radio emitting particles
and magnetic field of jets. 

Below we discuss a few large monitoring programs carried out at the Northern 
Sector of RATAN-600 having resulted in the SS433 light curves at 4--6 frequencies.

\section{Observations and processing}

A part of observations was carried out in the program of the
studies of variable X-ray sources in long-term cycles of the monitoring of
flare activity (Trushkin 2000, Trushkin et al. 2001a,b). A regular complex of
continuum spectrum radiometers was used. The receivers of frequencies 3.9, 
7.7, 11.2 and 21.7 GHz were equipped with cryogenic systems of closed loop
that depress the temperature of the first amplifier stages down to 15--20~K.
The radiometers of frequencies 0.96 and 2.3 GHz were equipped with low-noise 
transistor amplifiers. Besides, to avoid the narrow-band interference
from mobile communications and locators, the wide bands of these receivers 
were cut into 4 and 8 subchannels correspondingly. It should be
mentioned that from the midyear 2002 there have been no observations at
960 MHz because of a powerful interference generated by cellular
transmitters in a range of 910 MHz mounted near the RATAN-600.

The observation was fulfilled by the RATAN-600 Northern Sector or Southern 
Sector plus a Flat reflector every day at the source culmination, i.e. the
UT time of observations was changing during every cycle, whereas the local
sidereal time of observation was remaining the same. In an observation the flux from
a source was recorded at two--six frequency ranges. In spite of a bright
galactic background the telescope angular resolution is sufficient
to detect reliably SS443 on drift scans.

Table 1 lists observational sensitivities of the telescope RATAN-600 (a 
level 1$\sigma$ at the optimal smoothing of records) in a single transit of
the source through a fixed beam pattern. Unfortunately, because of frequent 
interferences the pointed sensitivity was not always achieved in real 
observations. Generally the antenna of the Southern Sector with the Flat
reflector had a noticeably lower system flux sensitivity because of higher antenna
noise temperature and smaller effective area in comparison to the antenna of
the Northern Sector. 

On November 6, 2003, the receiver at frequency 3.9~GHz was replaced by a
new high sensitive cryogenic two-channel radiometer at frequency 4.9~GHz.

Every day during all cycles the calibration sources 1345+12, 1850-01 or
2128+04 were observed. Their current fluxes were calibrated using the secondary
calibration sources: 3C286, NGC7027, DR21 and some others.
We suggest that the SS433 flux measurement error does not exceed 3--8{\%}
at all frequencies: 0.96, 2.3, 3.9, 11.2 and 21.7~GHz.

\begin{table*}
\caption[]{Flux densities of reference sources (Jy)}
\medskip
\begin{center}
\begin{tabular}{l|l|l|l|l|l|l|l|l}
\noalign{\smallskip}
\hline
Name           & Another    & \multicolumn{7}{|c}{Frequency (GHz)} \\
\cline{3-9}
  of source    & name &  0.96  &  ~2.3   &   ~3.9   &  ~4.9   &  ~7.7    &  11.2  &  21.7 \\
\hline
 1328+30 & 3C\,286          &  17.2  &  11.5  &   ~8.55  &  6.45   &  ~5.52  &   ~4.25 &  ~2.5 \\
 1345+12 &                  &  6.48  &  ~4.4  &   ~3.31  &  2.55   &  ~2.21  &   ~1.74 &  ~1.12\\
 1850-01 & HII              &        &   2.27 &    3.06  &  3.63   &   3.88  &    4.19 &   4.49\\
 2037+42 & DR\,21           &  ~5.0  &  12.1  &   17.4   & 19.9    &  21.7   &   21.7  &  20.7 \\
 2105+42 & NGC\,7027        &  ~0.9  &  2.64  &   ~4.9   &  5.54   &  ~6.33  &  ~6.03  &  ~5.9  \\
\hline
\end{tabular}
\end{center}
\end{table*}

The fluxes for secondary calibration sources were taken from a paper by Aliakberov et
al. (1985), which in turn were in the basic radio-astronomical
flux scale of reference sources by Baars et al. (1977) and with updated
flux measurements in a paper by Ott et al. (1994). The fluxes of reference
sources accepted for the present measurement cycle are given in Table~2.

The digital recording of data, preliminary processing and archiving were made by
the acquisition software packages created by P. Tsybulev and V.K. Kononov. The
processing of obtained records of target and reference sources was
made with a data-processing program ``{\it prat}''. Data processing
included the procedures of background and pulse interferences removal, the 
convolution with the beam pattern, and the Gauss analysis.

\subsection{Study of SS433 in the experiment ``Cold''}

Undertaken in 1980 a 100-day observational set on searching for microwave background
fluctuations  was carried out at a fixed antenna angle of
elevation 51${\degr}$ (i.e. at a declination chosen to observe every day SS433
Dec1950~=~+04${\degr}57'$) with an unmoved feed-cabin (Parijskij {\&}
Korol'kov 1986). This mode is the most optimal to study the fast flux 
variability of radio sources. Thus, dozens of instantaneous SS433 spectra in the
range from 3.9 to 31 cm were obtained in daily observations. In February--May
1980 SS433 did not show strong variability. The absence of bright flares 
allowed carrying out the Fourier analysis of the SS433 radio brightness with 
the purpose of searching for radio flux periodicity reflecting
intrinsic binary periodicities: 13-day orbital period or its half of 6.5 days.

A low detectable harmonic of 6.5~$\pm $~0.5 days appeared in a sample of the cycle
onset in February--March 1980, but later it was not detected. In the
13-day-average  observations at a wavelength of 2.08~cm there is also a detectable
harmonic of the 13-day period, but the data were not sufficient for final 
conclusion about the existence of periodicity in the quiescent radio emission
of SS433. 

Only on basis of many observations in 1997 and 1999 and due to a precise 
calibration by the flux density we managed to unambiguously detect a 
6.05-day modulation of the quiescent  radio emission of SS433 at all the six
frequencies (Trushkin et al. 2001).

\begin{figure*}
\centerline{\vbox{\psfig{figure=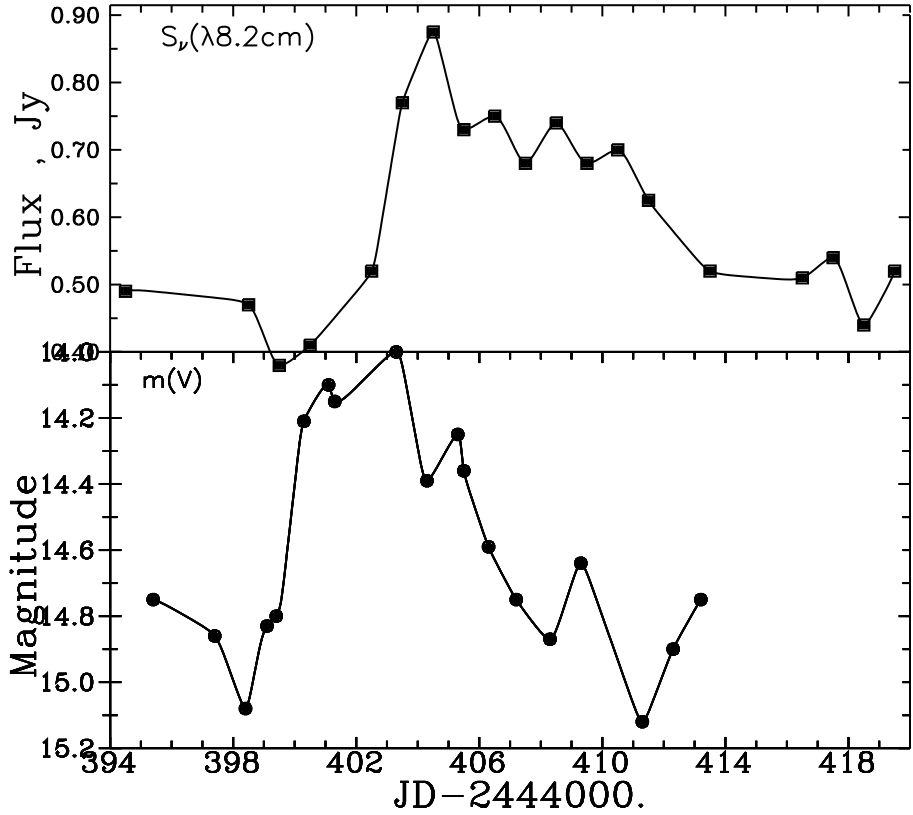,width=12cm,angle=-90}}}
\caption{%
The SS433 light curve in V band and the curve of flux variations
at 8.2~cm wavelength in June 1980.
}
\label{ssf01}
\end{figure*}

A faint flare that is seen the best at  8.2 cm wavelength correlates with
variations of the photometric flux (Gladyshev 1981). In Fig.\ref{ssf01} a typical
delay of the radio flare relative to the optical one and the general similarity
of light curves are clearly seen. Neizvestnyj et al. (1980) also made a
conclusion about the delay of radio flares. In observations of 1980 a
substantial linear polarization of SS433 radiation at a level of 5--10{\%} was first
detected. Hjellming and Johnston (1981) showed that the high linear
polarization is related to jet ejections, whereas the central core
($<0.1''$) is not polarized. On the average, the SS433 flux in the centimeter
range in the quiescent state of this cycle is fitted well by the formula
$$S_{\nu}{\rm (Jy)}=1.1\nu ^{ -0.6} {\rm(GHz).}$$

There are near 300 other observations of SS433 during other ``Cold'' sets from 1988
to 1994 that are not included in the paper.

\subsection{General behaviour of the SS433 radio emission}

We list all measurements carried out  with the radio telescope
RATAN-600 from December 1986 to January 2004 in  Table 3.
The first column of the table gives a modified Julian date of observation
 MJD~=~JD-2400000.5,
columns 5--8 give flux densities (in mJy) at  frequencies of 0.96, 2.3,
3.9, 7.7, 11.2 and 21.7~GHz. Note that from November 6 of 2003 the receiver at
3.9~GHz was replaced with a two-channel cryo-radiometer at a frequency
of 4.9~GHz. The absence of some numbers from these columns means that either
there was no observation or it failed because of one or another reason 
(atmospheric or electric interferences, a radiometer trouble, etc.). 
Column~9 gives the calendar date of observation {\mbox (ddmmyy)} and in columns 10 and
11 a spectral index and its error are given, they are determined from a
fitting of a 2--6-point spectrum by a single power law:
$S_{\nu}=S_{o}\nu^{\alpha}$. Summarily there were 940 observations of
SS433, and 4500 flux densities were measured.

The data of the flux density measurements are publicly-accessible as 
a web-program for plotting the SS433 spectra `on-line' on the web-server of
the CATS data base (Verkhodanov et al. 1997):
http://cats.sao.ru/cgi-bin/ss433.cgi. To get the spectrum plots and
fitting parameters directly on the monitor screen a user is only to choose a 
modified Julian (MJD) or calendar date of the SS433 observation.

\begin{figure*}
\centerline{\vbox{
\psfig{figure=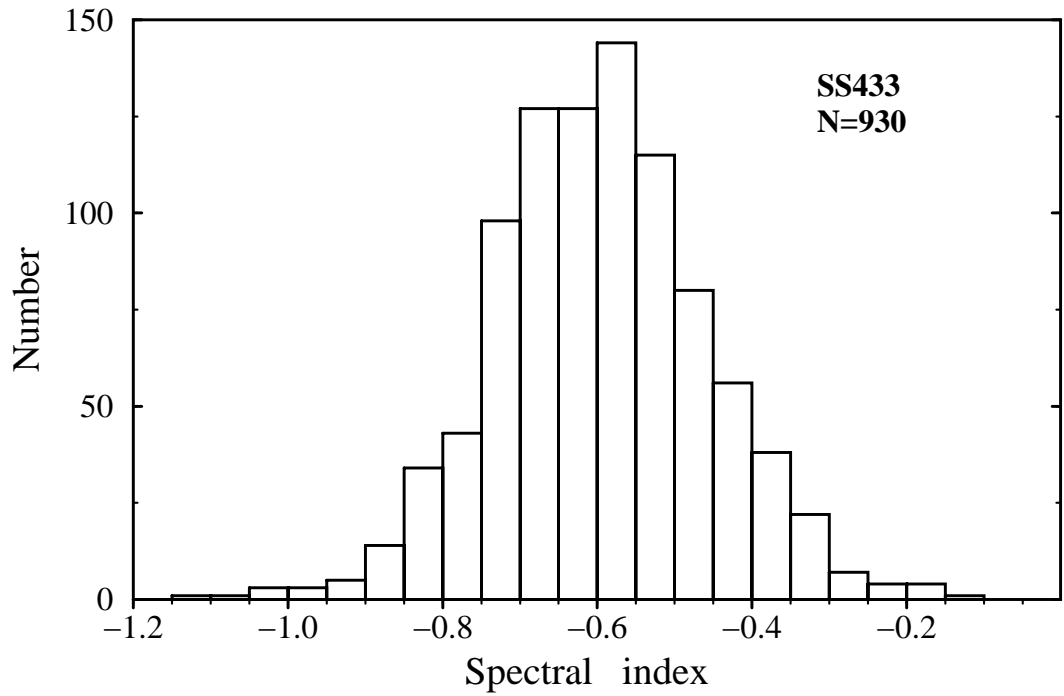,width=14cm,angle=-90}
}}
\caption{%
The bar chart of the SS433 spectral index distribution by
930 measurements wih RATAN-600.
}
\label{al_his}
\end{figure*}

Fig.\ref{al_his} shows a histogram of spectral index distribution by total data from Table 3.
The daily spectra were well fitted by a power law relation, a mean spectral
index remained the same, $-0.60\pm 0.14$ over 20 years, and the spectral index
determination error (column 11 in the table) of our measurements being
better than 0.1 on the average. It is intriguing that in the centimeter 
wavelength band the spectrum of SS433 is always optically thin with the
spectral index $\alpha $ varying from $-0.2$ to $-1.4$, i.e. there is no
case with a positive index or inverse spectrum. However, a reservation should be
made: we are dealing with an integral spectrum of SS433; if we subtract a 
``quasi-constant'' component of the flux, then the flares spectra can be 
inversion ones evidently because of the delay of flares at low frequencies.
The mean flux for $\sim $~600~ measurements at a frequency of 960 MHz is equal
to 1530 mJy. 

It should be noted that in our observations the spectral index measurement 
accuracy is higher than that in a two-frequency program of SS433 monitoring with
the three-element radio interferometer GBI in Green-Bank (USA). From
publicly-accessible data of GBI one can estimate an average spectral index 
to be $-0.70\pm 0.30$ by 15000 measurements at frequencies of 2.25 and 8.3
GHz. As was shown by the analysis of measurements fulfilled in the same 
time, this significant difference in the average spectral index is due to 
the fact that GBI fluxes at 8.3~GHz are underestimated by 10--15~{\%}, though
``our'' 2.3-GHz fluxes and GBI 2.25-GHz fluxes practically coincide, what 
resulted in a steeper average spectrum of SS433 by GBI data in comparison 
with what was obtained by our data. One can assume that the cause of a lower 
measured flux at the high frequency is a partial resolution of SS433 by the 
GBI interferometer with the base of 2.4~km. 

\subsection{Search for regularities in the radio flares of SS433}

In December 1986 -- March 1987 a long-term cycle of observations of SS433 
radio flares was carried out. Six rather powerful flares were detected at
wavelengths of 3.9, 7.6, 13 and 31.2~cm during this 90-day cycle. Fig.\ref{my86} shows
light curves at these frequencies. 

During 2--3 weeks of May 1987 an international multi-wave monitoring campaign
 was carried out on SS433 (Vermeulen et al. 1993a,b), when this object
was observed in coordination with several radio telescopes (MOST, VLA, NRAO 
GBI, RATAN-600, Bologna Cross and European VLBI) and many optical 
telescopes. 

\begin{figure*}
\centerline{\vbox{
\psfig{figure=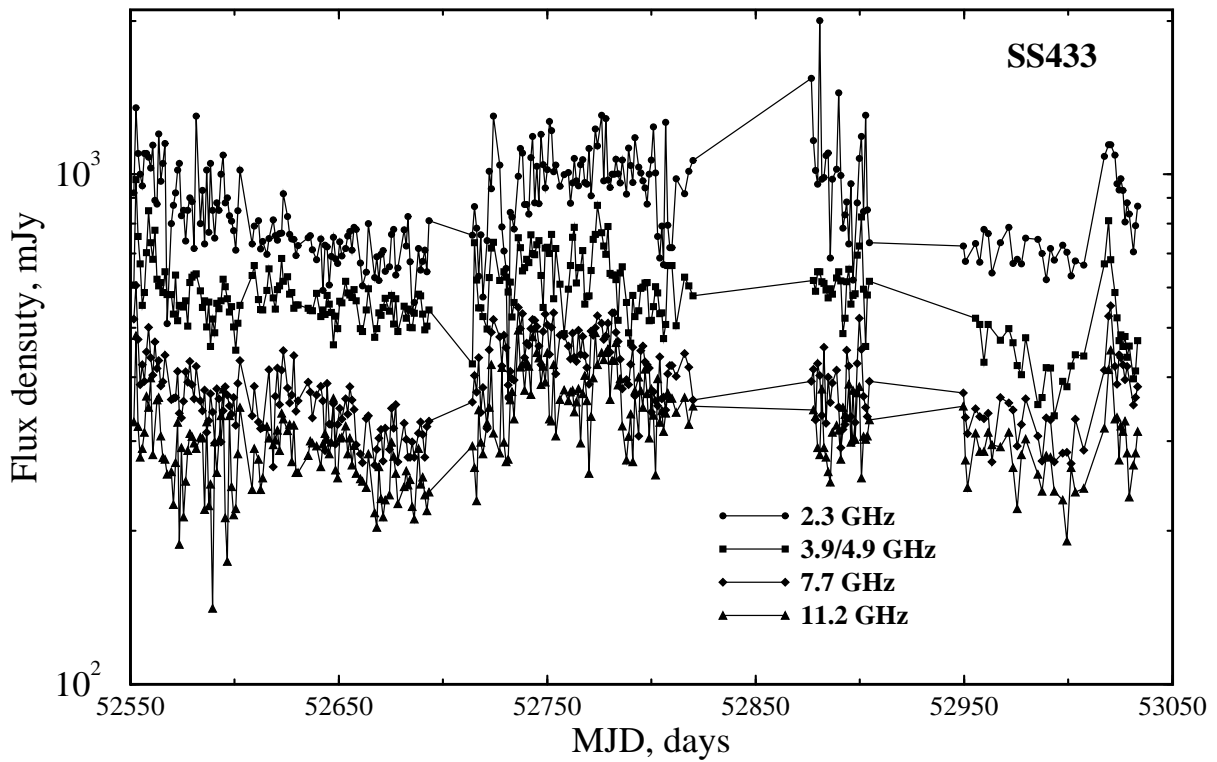,width=16cm,angle=-90}
}}
\caption{%
The long-term period of SS433 observation with RATAN-600
from October 2002 to January 2004.
}
\label{ss_my2}
\end{figure*}

\begin{figure*}
\centerline{\vbox{\psfig{figure=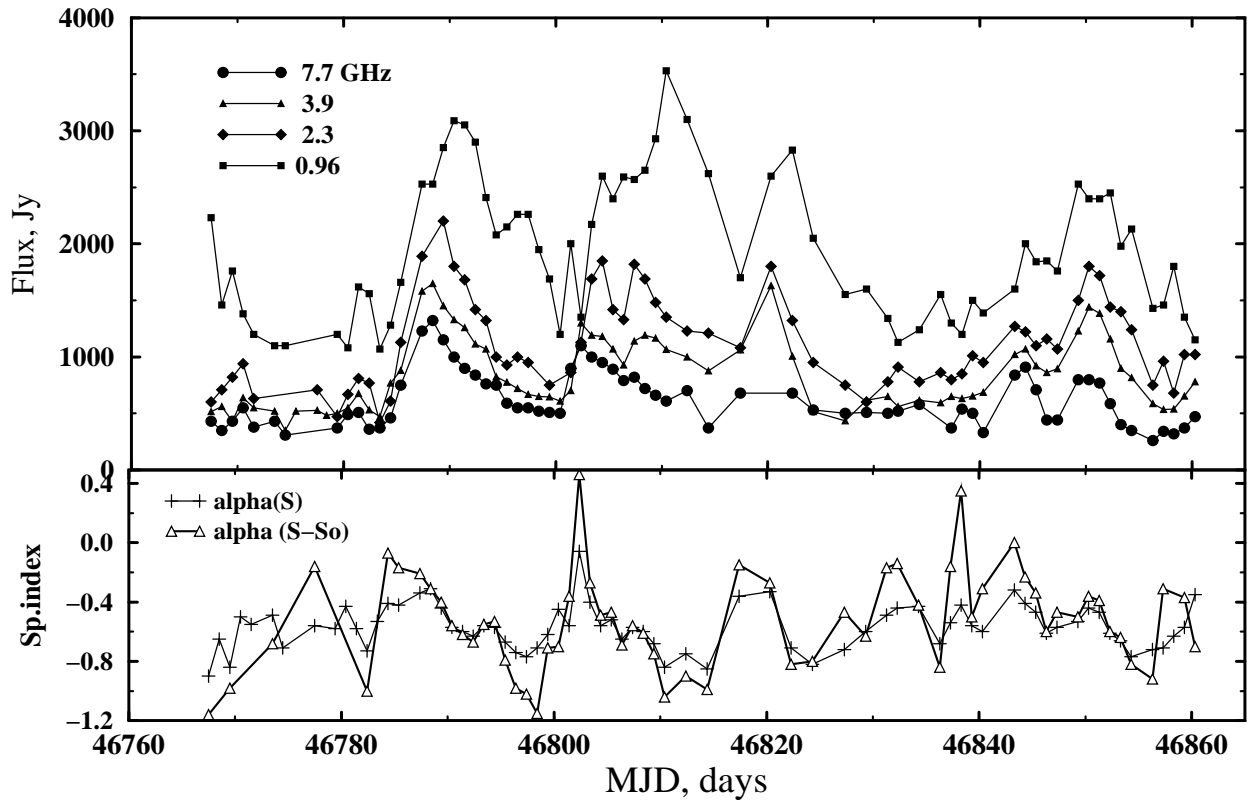,width=16.5cm,angle=-90}}}
\caption{%
The radio emission light curves of SS433 at wavelengths
3.9, 7.6, 8.2, 13 and 31~cm in December 1986 -- March 1987 and
variations  of the spectral index of the total radio emission and that
after subtraction of the quiescent flux levels.
}
\label{my86}
\end{figure*}

\begin{figure*}
\centerline{\vbox{\psfig{figure=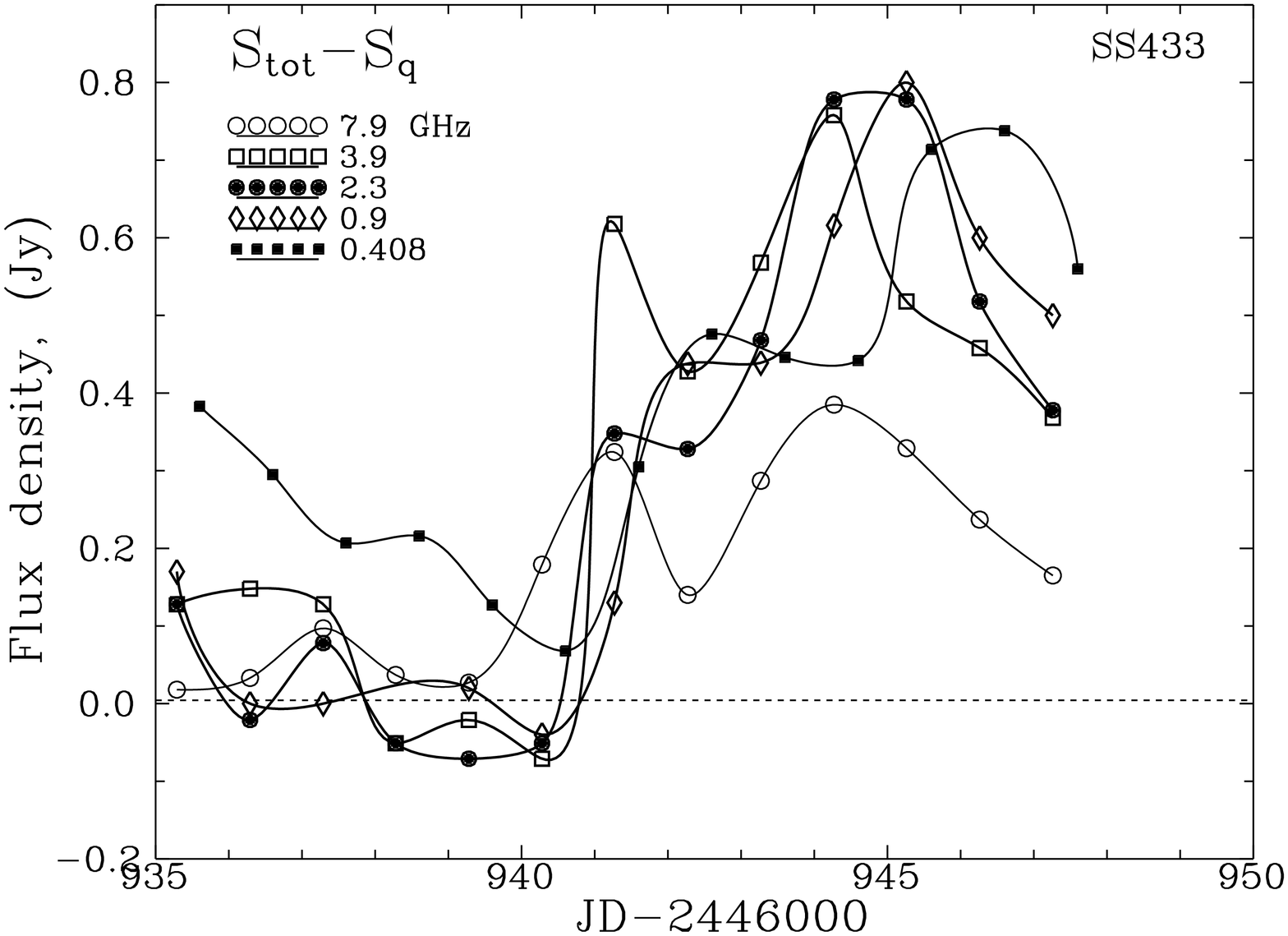,width=12cm,angle=-90}}}
\caption{%
The radio emission light curve of SS433 at different
frequencies in May 1987. The constant part of flux (S$_{\nu }$~=~1.1~$\nu 
^{ - 0.58})$ was removed. The data at 843 and 960~MHz and at 7700 and
8085~MHz were averaged.
}
\label{my87}
\end{figure*}

In Fig.\ref{my87} we present the curves of flux variability (S$_{tot}$~--~S$_{q})$
during 19--31 May 1987. It is seen that in this period we managed to record
two powerful flares, the second one being apparently related to the increase 
of flux in a separate component (a blob) of the jet in a so-called zone of
``brightening'' (Vermeulen et al. 1993a). The recent active mapping of
SS433 with an interferometer of the highest resolution {\mbox VLBA} (NRAO, USA)
undertaken in Summer 2003 during 42 days (Mioduszevski et al. 2003) have 
shown that the blobs do not all necessarily become brighter on the 
5$^{th}$--6$^{th}$ day of the travel from a center (50\,mas) what has been shown
by Vermeulen et al. (1993b). Such brightening can occur earlier or not
occur at all, and the brightening phenomenon can be related to selected 
azimuths of jets. Sometimes the blobs brightening can be very
strong, up to 30 times and occur at equal distances/times from XBs (subject 
to relativistic effects). The motion conditions of separate blobs change 
depending on the activity of so-called ``anomalous'' outflows across jets or 
near the plane of accreting disk of this XB. The velocity of these motions 
($\sim $~10000~km/sec), which was measured in this set for the first time, 
is so high that it necessarily impacts a lot the filling of a channel 
generated by the motion of jets along the 164-day precession trajectory.
Interestingly, the motions at velocities of 0.1--0.15\,c in the radio structures
of SS433 were also detected earlier (Stirling et al. 2002).

In Fig.\ref{ss_my2} we present the  radio emission light curves of SS433 during a
long-term observational set of duration almost one and a half year in 
2002--2004. A rather quite period during a year was broken by a relatively
powerful flare only in the end.

Here we should touch on a very interesting phenomenon --- a pre-flare flux
fall below an average quiescence level at all frequencies. The ``negative''
fluxes in Fig.\ref{my87} are caused by such a ``dip''. Such short
($\sim 1-2$\,days)
dips were observed in SS433 more than once, although not before all flares.
 The dips of the quiescent emission are estimated as $\sim 100-200$~mJy.

At least three such dips of the SS433 radio emission before the
flares in  MJD52213, 52325 and 52274 are seen in Fig.\ref{my86}. It
is interesting that between the first and the second flares the observations 
with the satellite RXTE in a range of 2--12~keV were fulfilled and in a number
of measurements the detection of strong X-ray flux variability has been 
gained in an interval about 3000 seconds (Kotani et al. 2002). In Fig.\ref{ss_my2}
the dip of SS433 radio emission before the flare in the interval
MJD52985--52995 is well noticeable.

Perhaps, the dips are related to the appearance of optically thick matter,
that resulted from heavy accretion prior to powerful flares (Waltman et al. 1994).
Then either optical depth of this matter decreased because of fast expansion
or a plasmon, which is responsible for the flare, ``resurfaces'' from this absorbing
matter and deletes the flux deficit. Since the dip  concerns the
quiescent component of
the SS433 flux, all environment of the binary system as well as jet
insides are supposedly rearranged. 

The phenomenon of dip  was observed only before the first flare
after a long quiescent period, but it is not improbable that we cannot see it
before all flares because of an overlap of flares. For example, in the flare 
of May--June 1996 (Fig.\ref{f96}) there are no visible flux fall-down at all
frequencies.

\begin{figure*}
\centerline{\vbox{
\psfig{figure=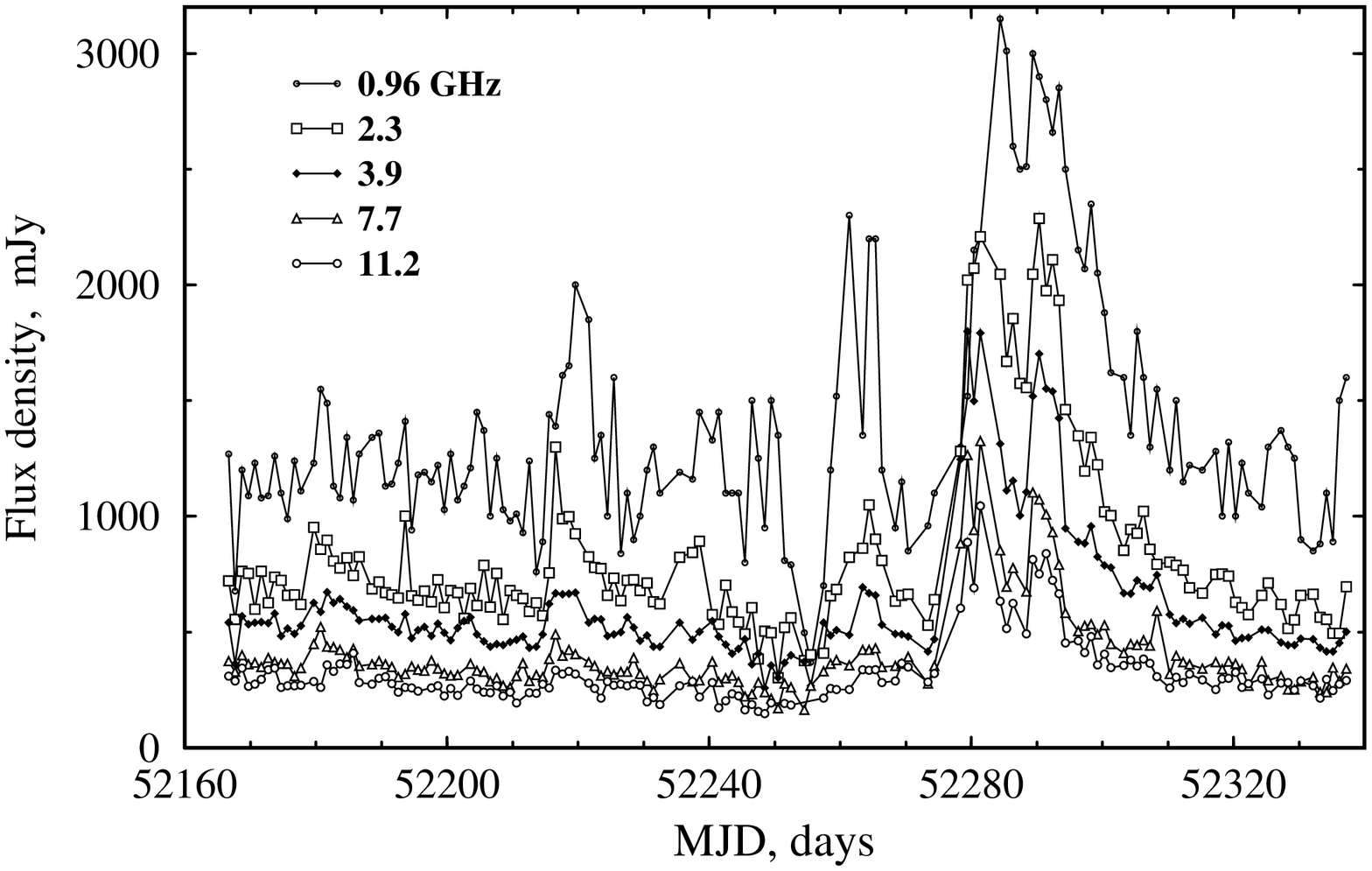,width=16cm,angle=-90}
}}
\caption{%
Long-term observations of SS433 with RATAN-600 from
September 2002 till March 2003.
}
\label{ss_my1}
\end{figure*}

Finally, it is noteworthy that an analogous effect is pronounced more 
strongly in another microquasar Cyg~X-3, where the flux dip
duration correlates with the power of subsequent flare and the lowest (at 
the detection threshold) level of hard X-ray flux (Trushkin 1998).

Using all SS433 observations carried out with RATAN-600 including 
reprocessed observations from a paper by Neizvestnyj et al. (1980) we 
obtained average characteristics of radio flares. 

The flares last 3--15 days, the flux density variation can reach 0.5--0.8~Jy at a
frequency of 7.7~GHz and 2--3~Jy at 960~MHz. The main flare parameters were
analyzed: the dependence of the flare maximal flux $\Delta $S$_{m}$ and its 
time $\Delta $t$_{m}$ on frequency. It turned out that these relations are 
well fitted by power laws with indices indicated for each flare. Average
over 9 flares parameters of these relations are below.
$$\Delta{t_m}=\Delta\,t\cdot\nu^{m},~m=-(0.20-1.0), ~\overline{\Delta{t_m}}=5.1\cdot\nu^{-0.43}$$
$$\Delta{S_\nu}=\Delta\,S\cdot\nu^{n}, n=-(0.15-0.9),~\overline{\Delta{S_\nu}}=1.3\cdot\nu^{-0.46},$$
where in the last formulae the flux is in Jy and the frequency is in GHz.

\begin{figure*}
\centerline{\vbox{\psfig{figure=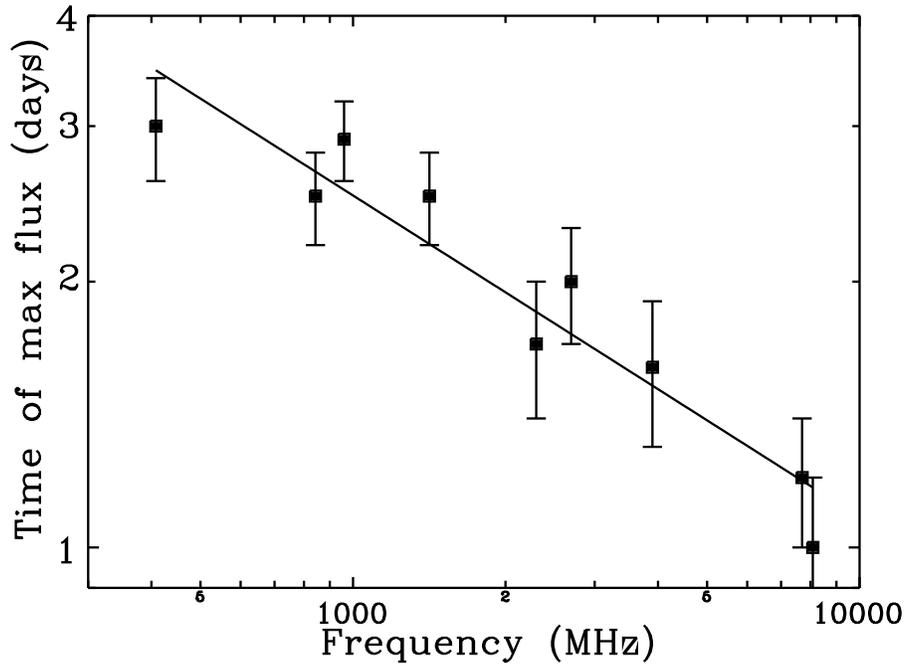,width=12.0cm,angle=-90}}}
\caption{%
The change of maximal flux times $\Delta $t$_{m}$ with
frequency of the SS433 flare in May 1987. Time is counted off an assumed 
flare onset on May 24 (JD~2446940.0). The line slope corresponds to a power
law with an index of m~=~--0.36.
}
\label{D87}
\end{figure*}

\begin{figure*}
\centerline{\vbox{\psfig{figure=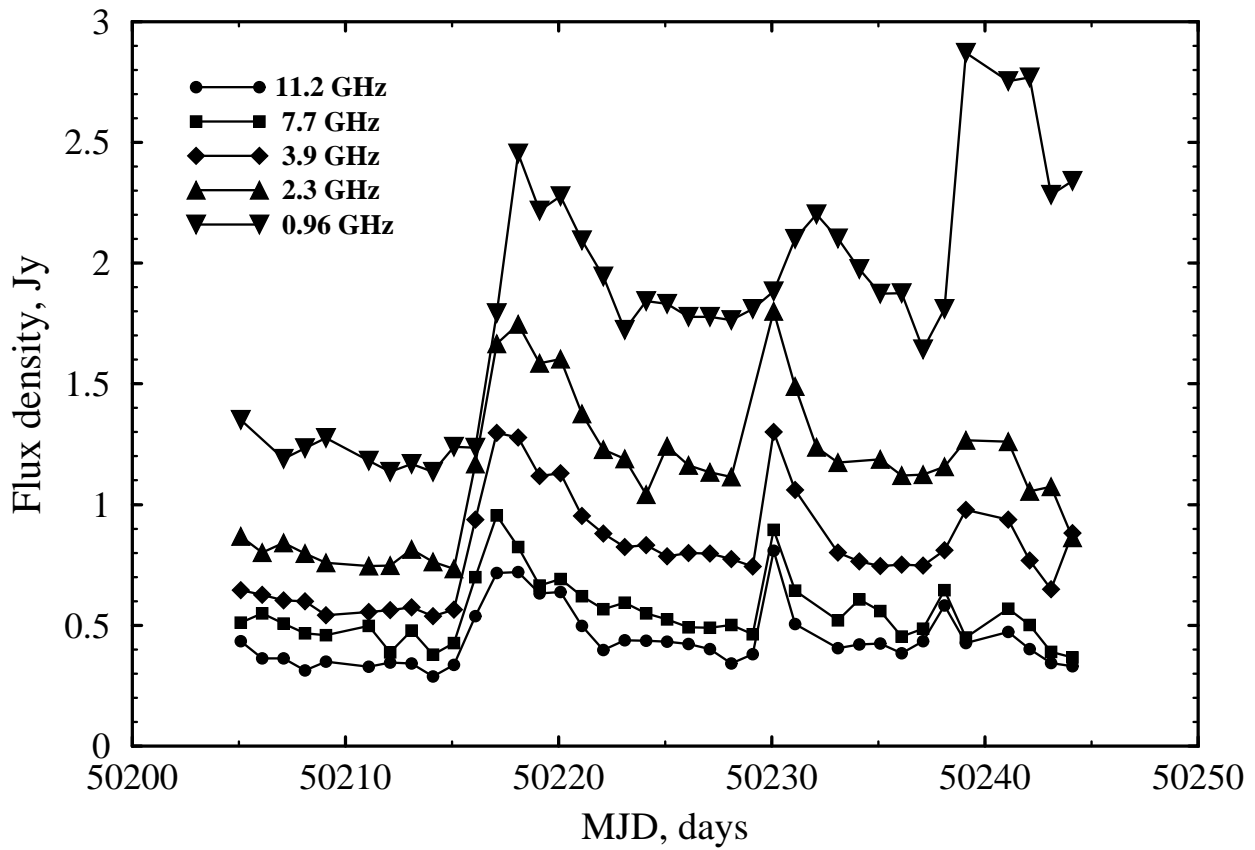,width=16.5cm,angle=-90}}}
\caption{%
The light curves during the flares of SS433 in May 1996.
}
\label{f96}
\end{figure*}

What the above relations are valid during long period of time is the most 
important property of radio flare. The flare increase is always steeper than 
its fading that indicates a different size of emission region at flare
onset and end. Fig.\ref{D87} shows how reliably the power law for $\Delta $t$_{m}$
in the first flare of May 1987 was determined. 

A noticeable flare delay towards lower frequencies and step-by-step
after-flare increase of the spectrum steepness was registered in a series of 
flares in May--June 1996. A rather different spectral and temporal character
of successive flares also engages our attention. 

\begin{figure*}
\centerline{\vbox{\psfig{figure=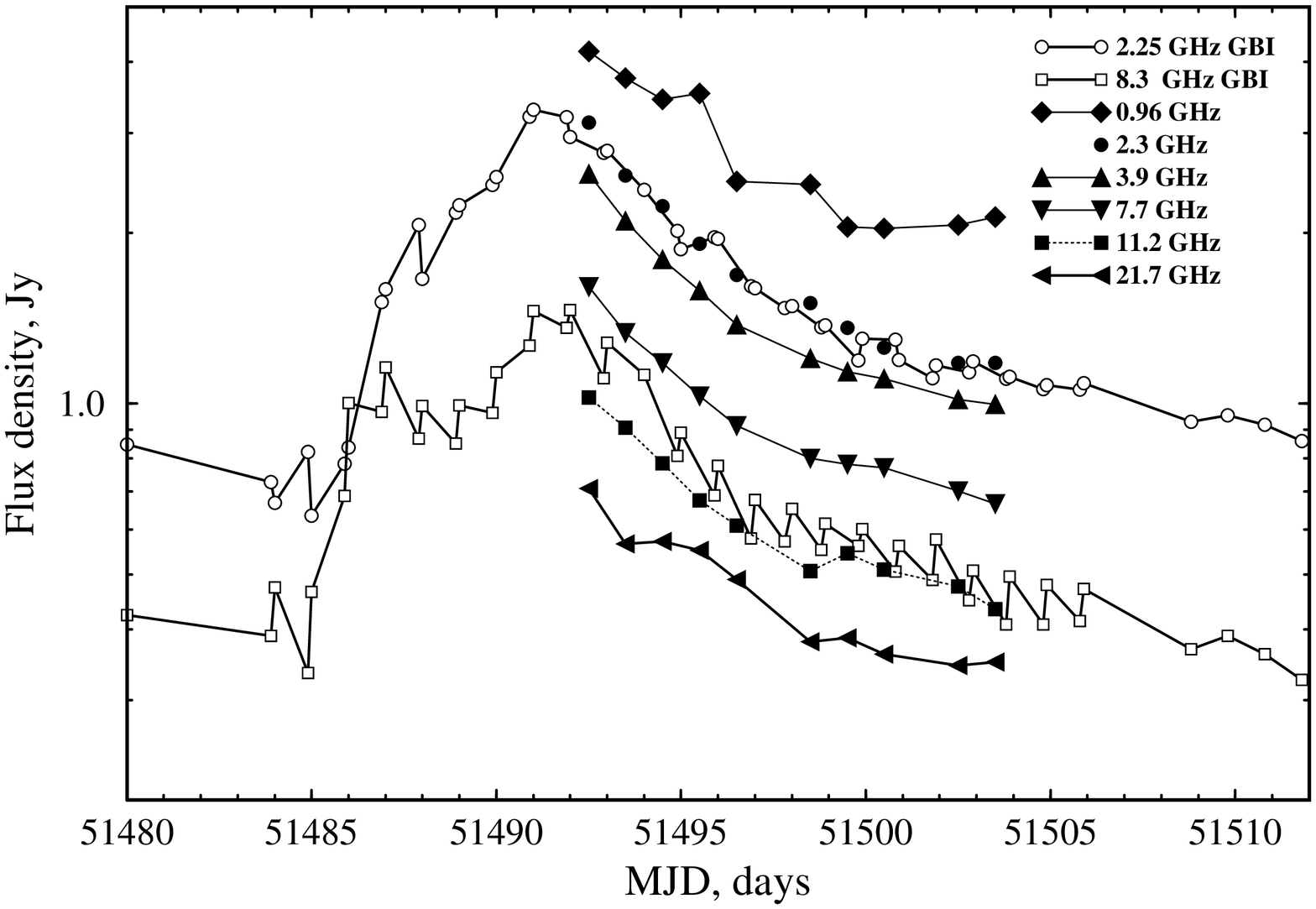,width=14cm,angle=-90}}}
\caption{%
The light curves of a very bright flare of SS433 in
November 1999 by GBI data at 2.3 and 8.3 GHz and by RATAN-600 data at 
different frequencies. 
}
\label{99n}
\end{figure*}

Fig.\ref{99n} shows the light curves of a very bright flare of SS433 in November
1999. At 2.25 and 2.3 GHz the data of GBI and RATAN-600 coincide, but at
 8 GHz a noticeable difference is seen. In GBI data the fluxes at 8.3
GHz are underestimated by 10--15{\%} resulting in a steeper average GBI spectrum
of SS433 than it follows from our data. The duration of this flare 
indirectly indicates that the angular size of the source can be comparable 
to a synthesized beam pattern of the 2.4-km interferometer GBI, what results 
in a noticeable underestimation of the flux density at 8~GHz.

\begin{figure*}
\centerline{\vbox{\psfig{figure=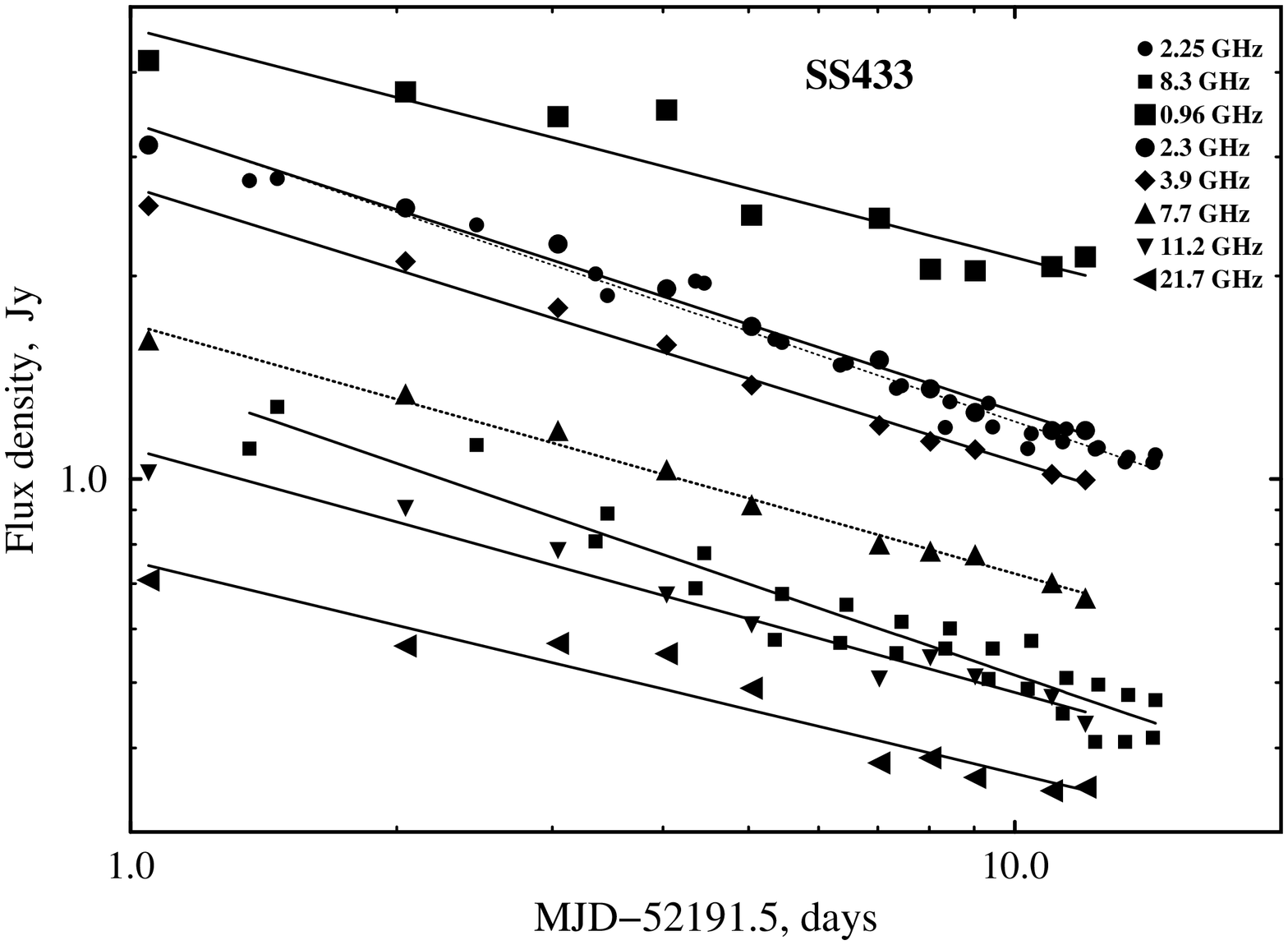,width=14cm,angle=-90}}}
\caption{%
The light curves of the SS433 flare in November 1999 after its maximum which are
well approximated by power laws at different frequencies.
}
\label{99nlog}
\end{figure*}

If we count off the start of the SS433 flare fading in November 1999 from the
moment MJD5149.5, then we can try to determine the fading law. The analysis
shows that it is preferable to choose a power law. The light curves at all 
frequencies show fading according to the law $\sim $~t$^{ \mbox{--}0.4}$ with a
correlation factor exceeding 0.96 (Fig.\ref{99nlog}).

\begin{figure*}
\centerline{\vbox{\psfig{figure=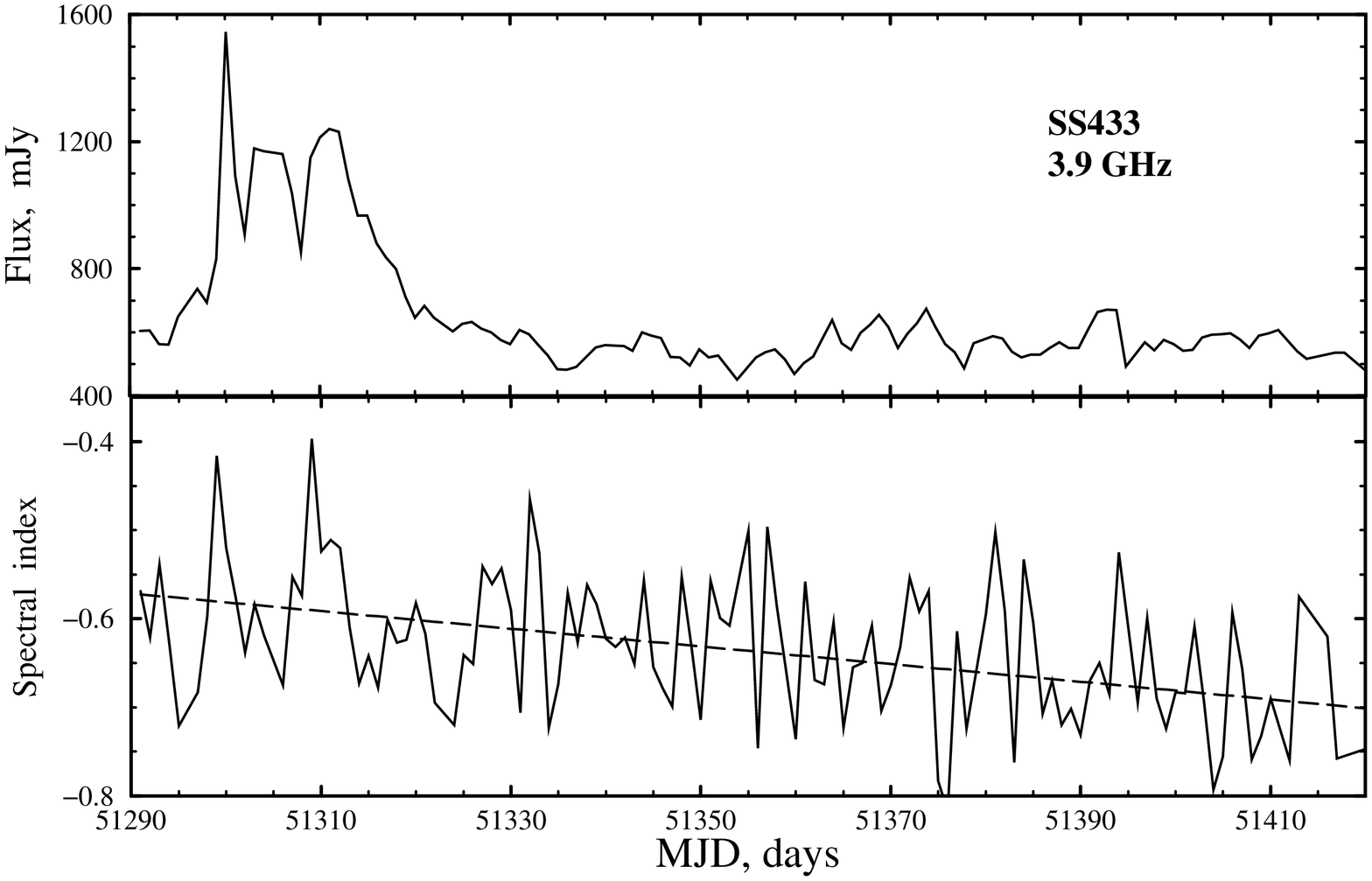,width=16.5cm,angle=-90}}}
\caption{%
The SS433 light curve at a frequency of 3.9~GHz (above) and
variations of the spectral index (below) in December 1999 -- March 2000.
}
\label{al99}
\end{figure*}

\begin{figure*}
\centerline{\vbox{\psfig{figure=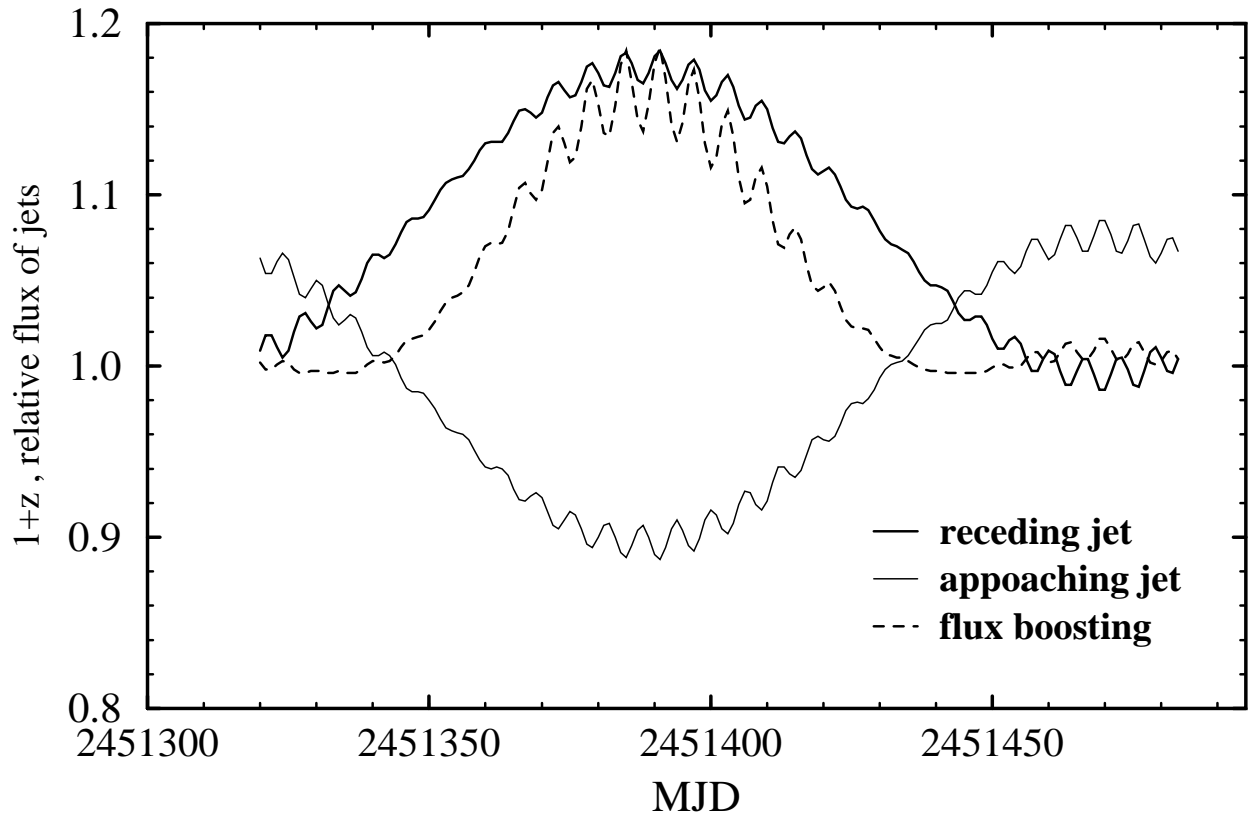,width=16.5cm,angle=-90}}}
\caption{%
Predicted by the kinematic model Doppler shifts
during the May--July 1999 and relative flux variations of SS433, appeared due to
the relativistic Doppler boosting.
}
\label{fig13}
\end{figure*}

Fig.\ref{al99} shows one of the obtained curves at a frequency of 3.9~GHz in long-term
daily observations of SS433 in December 1999 -- March 2000. Below in the
diagram the alteration of spectral index is presented. An abrupt change of 
the index during two flares and a step-by-step after-flare increasing of the 
spectrum steepness due to more evident flux fall at high frequencies $>8$~GHz
is easily perceptible. It is in this after-flare part of the light curves
where we detected a 6.05-day modulation with amplitude about 10{\%} 
(Trushkin et al. 2001) possibly due to relativistic boosting of
emission from jets in proportion to their ``nodding'' movements caused by 
the nutation of accreting disk and, correspondingly, the jets with a typical 
doubled frequency of orbital (13.08d) and precession motion. 
In order to illustrate such possibillity we plotted in Fig.\ref{fig13}
 the predicted by kinematic
model Doppler shifts around our set of 1999, allowed for the nodding
motion (Vermeulen 1989). The flux variations more than
five per cent are clearly seen in the sum of two fluxes from the receding
and approaching jets, received for case of discrete blobs in jets.
Additional effect could be obtained if the ``lifetime''  of these blobs is
compatible with the 6-day period.  The Fourier spectrum of this model curve
show the presence of the 6.05-day harmonic.

\section{Conclusion}

This paper compile all observational data over many sets from December
1986 to January 2004 containing 940 observations of SS433 at two--six
frequencies simultaneously. In all more than 4500 measurements of flux 
density at frequencies from 960 to 21700~MHz are presented. The average 
spectral index by all data is equal to --0.60~$\pm $~0.14 with an average 
measurement error of 0.09 at an average flux density of 1.5~Jy at 960~MHz. In the
indicated frequency range there were no cases of the spectrum inversion, 
when the spectral index would be positive. Thus, if there exists such a 
stage of the SS433 spectrum evolution, it is essentially shorter than one 
day not to get in such a moment during almost 1000 observations. 

It was demonstrated that there are some earlier unknown regularities of the 
SS433 activity in the measured light curves. By the data on many flares
the delay of flare maximum towards lower observational frequencies was detected.
On the other side, the value of this maximum falls with frequency according 
to a power law. The flares fading with time also follows a power law. The 
indices of these relations change from flare to flare, but median values of
indices coincide astonishingly well. They are equal to --0.4~$\pm $~0.1. It
is evidently an indication that the integral emission of the SS433 jets
is due to similar properties of synchrotron radiation of the
blobs--plasmons moving separately. In the end it is rigidly related with
processes inside this XR and jets on the whole. From this point of
view the models with inner shock waves propagating  and
increasing radio emission along the jet could be more preferable than
the models in which the evolutions of separate radio components, blobs, are 
independent from each other. 

The accumulated data of the flux density measurements are presented as 
publicly-accessible Web-programs for constructing the SS433 spectra in the 
Home Page of the data base CATS: http://cats.sao.ru/cgi-bin/ss433.cgi.

The complete data of all measurements carried out with RATAN-600 are 
presented in Table~3. 

\begin{acknowledgements}
The work was carried out with the support of the RFBR grant
02-02-17439 in 2002-2004.
The authors express their sincere gratitude to the RATAN-600
maintenance service for the good work of antenna and
to our colleagues for the help in observations and
processing of data. Also the authors are grateful to the
GBI interferometer staff for a possibility of using
their observational data of SS433.
\end{acknowledgements}

\clearpage
\begin{onecolumn}
\topcaption{\large Flux density measurements of SS\,433}
\tabletail{\hline}
\tablefirsthead{\hline
~ MJD     & \multicolumn{6}{|c|}{Flux density in mJy at a frequency in GHz }        & Date   & Sp.   & Index  \\
~days~~~~ & \multicolumn{6}{|c|}{~~0.96~~~~~~2.3~~~~~~~~3.9~~~~~~~~7.7~~~~~~~11.2~~~~~~~21.7~~~}  &  ddmmyy& Index &  error    \\
\hline
~~~~(1)~~~&~~~(2)~~&~~~(3)~~~&~~~(4)~~~&~~~(5)~~~&~~~(6)~~~&~~~(7)~~~&~~(8)~~~&~~(9)~ &~~~(10)  \\
\hline}
\tablehead{\\
\hline
~~~~(1)~~~&~~~(2)~~&~~~(3)~~~&~~~(4)~~~&~~~(5)~~~&~~~(6)~~~&~~~(7)~~~&~~~(8)~&~~(9)~ &~~~(10)  \\
\hline}
\begin{supertabular}{lrrrrrrrrr}
46767.48  &2230     &  600     &  515    &   430     &          &      &031286& $-$0.78& 0.24 \\
46768.48  &1460     &  710     &  560    &   350     &          &      &041286& $-$0.67& 0.05 \\
46769.48  &1760     &  820     &  430    &   430     &          &      &051286& $-$0.72& 0.17 \\
46770.47  &1380     &  940     &  640    &   550     &          &      &061286& $-$0.46& 0.06 \\
46771.47  &1200     &  630     &  550    &   380     &          &      &071286& $-$0.54& 0.06 \\
46773.46  &1100     &          &  520    &   430     &          &      &091286& $-$0.46& 0.06 \\
46774.46  &1100     &          &  350    &   310     &          &      &101286& $-$0.64& 0.16 \\
46775.46  &         &          &  518    &           &          &      &111286&        &      \\
\end{supertabular}
\end{onecolumn}

\end{document}